# ФИЛОСОФСКИЙ ФИКЦИОНАЛИЗМ И ПРОБЛЕМА ИСКУССТВЕННОГО ИНТЕЛЛЕКТА*

*Куликов Сергей Борисович,*

декан факультета общеуниверситетских дисциплин,
доктор философских наук, доцент,

Томский государственный педагогический университет,
Томск, Россия

*kulikovsb@tspu.edu.ru*

**Аннотация.** Интерпретация искусственного интеллекта с позиций философского фикционализма позволила обсудить условия технологизации образования как процесса усиления уровня дискурсивной силы мышления учащихся. Искусственный интеллект получил расширительное истолкование как воплощение литературного образа, который не имеет однозначной связи с областью логических формализмов и математических алгоритмов действия и общения. Автор применил методы, свойственные семиотическому подходу, предложенному Борисом Успенским и Юрием Лотманом. Удалось подвергнуть критике современные версии технологизации образования, ведущие к приданию абсолютной ценности информационно-телекоммуникационным технологиям. Рост методологической культуры учащихся, оформленный в виде функционального словообразования новых средств описания изучаемых предметов, был связан с развитием дискурсивной силы мышления. Результаты исследования открыли новые перспективы в понимании места и роли искусственного интеллекта в рамках образовательной практики.

**Ключевые слова:** искусственный интеллект; фикция; фикционализм; ценности; образование; информационно-телекоммуникационные технологии; функциональный подход к словообразованию; искусственные языки; методологическая культура.

# PHILOSOPHICAL FICTIONALISM AND PROBLEM OF ARTIFICIAL INTELLIGENCE

*Kulikov Sergey B.,*
*Dean of University-wide Faculty, Doctor of Philosophical Science, Associate Professor,*

*Tomsk State Pedagogical University,
Tomsk, Russia.*

*kulikovsb@tspu.edu.ru*

**Abstract.** Interpretation of artificial intelligence on the base of the *philosophical fictionalism* allowed discussing the educational technologies in the context of conditions for strengthening of discursive power of pupils' thought. The artificial intelligence received broad interpretation as an embodiment of a literary image, which did not have unambiguous link with the area of logical formalisms and mathematical algorithms of action and communication. The author applied the methods peculiar to the semiotics approach, offered by Boris Uspensky and Yury Lotman. It was succeeded to criticize the modern versions of educational technologies conducting to giving of absolute value for information and telecommunication technologies. Growth of pupils' methodological culture, described on the base of functional approach to word formation of new means for the description of the studied subjects, was related to the development of discursive power of thought. As a result, the research opened new prospects in understanding of a place and role of artificial intelligence within educational practice.

**Keywords:** artificial intelligence, fiction, fictionalism, values, education, information and telecommunication technologies, functional approach to word formation, artificial languages, methodological culture.

**Введение**

Значительное место в исследованиях информационных технологий и киберпространства способно занять специфическое направление изысканий под названием «философский фикционализм». Корни философского фикционализма, который все-таки не стоит смешивать с фикционализмом математическим, восходят к идеям Ганса Файхингера [1]. Исходно фикционализм представал в виде учения о фикциях как вымышленных, но полезных науч-ных абстракциях («эфир», «сила» и т.д.). В современных условиях фикционализм суть одно из бурно развивающихся направлений, которое представлено в основном в зарубежной литературе и довольно слабо отражено в литературе отечественной. Причем современный фикционализм, если следовать логике работ последнего десятилетия, как это предлагает Фиора Салис [2], довольно сильно отошел от своих исходных предпосылок. Направление отхода во многом обусловлено лингвистическим нюансом, возникающим при попытке перевести слово fiction на другие языки буквально как «фикция», т.е. обман, иллюзия. Именно в таком смысле иногда говорят о фикционализме в рамках теории информации [3]. Но в английском языке fiction – это еще и беллетристика, художественная литература, и потому под фикционализмом понимают совокупность исследовательских установок, ядром которых является идея возможности изучения вымышленных объектов, т.е. таких объектов, которые не являются буквальной правдой. В дальнейшем, дабы избегать путаницы, мы используем кальку с английского для имени направления («фикционализм», «фикционалистский»), слово же fiction дается по-английски.

Реализация философско-фикционалистских наработок в контексте проблематики информационных технологий и киберпространства видится возможной в свете некоторых необходимых черт самого объекта исследования. Во-первых, информационные технологии и киберпространство, населенное роботами, шагнули в повседневную жизнь со страниц научно-фантастических романов. Все это можно напрямую воспринять в качестве разновидности воплощения fiction. Во-вторых, информационно-телекоммуникационные сети и автоматизированные системы как продукты информационных технологий по необходимости разделены на совокупность мест, имеющих зачастую вымышленные названия. Примером здесь выступает знаменитая интернет-энциклопедия «Луркморье» (Lurkmore), пользователи которой склонны принимать вымышленные имена и применять маски и псевдонимы.

Возникает вполне правомерный вопрос, в каком именно смысле фикционализм привлекается в этой статье и какой вклад автор надеется внести в проблематику. Привлекается фикционализм, затронутый в первом из указанных значений, причем привлекается для анализа чрезвычайно сложной темы, а именно искусственного интеллекта. Под интеллектом, как это представил Шон Стил, обычно разумеют так называемую «дискурсивную силу мышления» [4]. Данной силой может быть наделен как человек, так и его творения: компьютерные системы, роботы и другие. По понятным соображениям, в контексте данной статьи искусственный интеллект понимается как вид художественной литературы, точнее, он предстает в виде особой семиотической системы. Оправданность такой интерпретации обоснована в одном из специальных разделов статьи. Статья ориентируется на оценку перспектив применения «литературного» интеллекта в практическом плане. Именно обсуждение прагматических аспектов проблематики видится наиболее ценным результатом, который надеялся получить автор в ходе предпринятого исследования.

**Цель работы** – выполнить реконструкцию искусственного интеллекта в качестве особой литературной (семиотической) системы.

Материалы исследования включают результаты обсуждений возможности фикционалистского истолкования искусственного интеллекта. Интерпретация этих результатов позволила сформировать специальный методологический раздел статьи. Варианты практической реализации искусственного интеллекта в сфере образования послужили материалом для анализа, итоги которого представлены во второй части работы.

**Результаты исследования и их обсуждение**

*1. Методологическая база интерпретации искусственного интеллекта в виде fiction*

Необходимость прояснения методик представления искусственного интеллекта имеет как минимум два аспекта актуальности, а именно онтологический и технологический. Особенно важен онтологический аспект, ибо в современных условиях остается не до конца ясным, что же в конечном итоге имеется в виду в рамках словосочетания «искусственный интеллект». Тем более не ясно, что подразумевается в границах высказываний «искусственный интеллект есть реальность», «искусственный интеллект есть вымысел (fiction)» и в других утверждениях. В технологическом плане, имеющем достаточно узкий вид практического воплощения чего-то, что можно назвать «искусственный интеллект», остается не до конца понятным, как именно это «что-то» реализовать в полном объеме. Имеющиеся варианты реализации представляют собой алгоритмы поведения персонажей в компьютерных играх, либо же программы, обслуживающие работу автоматизированных и полуавтоматизированных устройств, допустим, конвейеров по сборке автомобилей, компьютеризированных систем по подготовке отчетности и других. Все это несколько отличается от того, что диктует воображение писателей-фантастов и сценаристов кинофильмов, изображающих искусственный интеллект как реальный субъект социальных отношений, который действует то в виде дружественной/недружественной человеку компьютерной системы (например, «Скайнет» в серии фильмов «Терминатор», «Красная королева» в зомби-хорроре «Обитель зла» и т.д.), то в виде также дружественных/недружественных человеку сообщества роботов. Мир, созданный воображением писателей и сценаристов, кажется эмоционально более привлекательным, нежели довольно скучный мир автоматизированных производств и офисных мероприятий.

Читателям книг и зрителям кинофильмов свойственно обманываться, принимая изображение искусственного интеллекта за его подлинную суть. Очень часто происходит смешение интеллекта, т.е. умения рассуждать и принимать решения на основе этих рассуждений, а также сознания и самосознания как способностей понимать чьи-либо действия, в особенности уметь отличить ошибочный ход мысли от верного хода, инновационный поворот от традиционного представления и многое другое. Как известно, компьютерные системы не ошибаются, а дают сбои; они не могут распознать в чем-то новом «хорошо забытое старое», а действуют в целом методом исключения. Другими словами, страхи перед искусственным интеллектом или же, напротив, чрезмерно позитивные ожидания от него обычно сопровождаются наделением простой рассудительной способности признаками умения оценивать результаты своих действий. Но «уметь рассуждать» и «уметь оценивать» – это все-таки не тождественные характеристики.

Объяснение природы интеллекта как такового все еще вызывает затруднения в современной науке. Научная мысль не предлагает достаточно строгих критериев ни для определения феномена «искусственный интеллект», ни для разведения интеллекта искусственного и интеллекта естественного. Так, развитие наиболее многообещающих в данном отношении когнитивистских иссле-дований, начатых в XX веке Ли Кронбахом [5] и подхваченных в современности Эрлом Хантом [6], рождает больше вопросов, нежели предлагает ответов. Во всяком случае, граница между искусственным и естественным интеллектом, которая в середине XX века казалась довольно четкой, все более стирается. Различение машинных и натуральных типов решения задач уже не кажется интуитивно ясным. Причем такое утверждение справедливо даже в свете наличия шахматных компьютеров, которые долгое время уступали шахматистам-людям, но в результате начали их обыгрывать. Победы компьютеров над людьми достигались не за счет некоей сверхчеловеческой логики, но благодаря возможности максимально быстро обрабатывать опыт имеющихся шахматных поединков, плюс в силу психофизиологических особенностей человеческого организма, склонности людей ошибаться под влиянием усталости. Так, память шахматного компьютера «Deep Blue», который обыграл Гарри Каспарова в 1997 году, включала несколько миллионов шахматных партий, но это были партии, сыгранные исходно людьми и только потом воспроизведенные машиной [7]. Компьютер выступил сверхскоростной эмуляцией шахматиста-человека, не добавив существенно нового в содержание шахматных партий. Новаторами шахматного искусства остаются люди. Во всяком случае, нет данных, чтобы соревнования, проводившиеся между компьютерными системами и людьми, либо же поединки только между компьютерными системами внесли существенный вклад в шахматы. Кроме того, не стоит сбрасывать со счетов, что компьютера-победителя в 1997 году программировали также люди, а не некие иные существа, имеющие нечеловеческий интеллект. Принципиально трудно представить, как вообще возможно, чтобы нечеловеческий интеллект сам создал для себя программу поведения и общения, отработал ее и реализовал в отношении людей. А ведь только эта ситуация могла бы служить примером проявления «настоящего» искусственного интеллекта.

Проблематический характер проведения границ между искусственным и естественным интеллектом особенно ярко отражен в работах некоторых философов последней трети XX века. Так, Сол Крипке [8: 144–155] и Ричард Рорти [9: 70–86], разбирая феномен боли и, главное, стараясь понять, как боль может быть адекватно выражена, описана и передана в процессе ком-муникации, обсудили два образа инопланетной расы. Первый образ включал описание существ, которые сводят все богатство психофизиологических переживаний к методикам стимуляции «С-волокон». Второй образ подразумевал возможность цивилизации, построенной существами, которые не обладают идеей ума (есть ли у них вместе с тем и сам ум, осталось под большим вопросом). В итоге были предложены модели возможного поведения, которые описывали субъектов, имеющих альтернативные чувства и интеллект относительно чувств и интеллекта человека. Причем подчеркивалось, что принципиально затруднительно найти общие способы интерпретации сообщений о процессах ощущения и размышления в рамках нечеловеческой и человеческой моделей поведения.

На правомерный вопрос, является ли так называемое «нечеловеческое поведение и общение» примером работы искусственного интеллекта, притом, что атрибут «нечеловеческий» относится к инопланетным расам, развивавшимся естественным путем на другой планете, можно ответить следующим образом. В контексте построения моделей, описывающих поведение и общение людей, атрибуты нечеловеческого играют ту же роль, которую играют атрибуты искусственного интеллекта в отношении интеллекта естественного. Отобразить это можно, представив три субъекта, обозначенные как S, S′ и S″. Причем S – это обладатель естественных (человеческих) качеств, S′ – это обладатель искусственных качеств, а S″ – это обладатель естественных нечеловеческих качеств. Знаком ≠ будем обозначать неравенство как нетождественность во всех отношениях, соответственно, равенство или тождественность во всех отношениях обозначим как =. Обобщим идеи Рорти и Крипке, допустивших, что S≠S′, а также S≠S″ в отношении чувств и ума. Другими словами, субъект S′, руководствующийся в частном случае искусственным интеллектом, и субъект S″, руководствующийся в том же отношении естественным, но нечеловеческим интеллектом, взаимно отличаются в общем случае от субъекта S как носителя человеческих качеств и свойств. Тогда получается, что S=S, а значит и S′=S″. Поэтому наличием признака естественности в отношении носителя нечеловеческого интеллекта можно пренебречь и считать его вариантом представления искусственного интеллекта, как альтернативы естественному человеческому интеллекту.

Предложенный подход не лишен слабых мест, во всяком случае, что касается фиксации сходства двух субъектов в отношении общих качеств в рамках перехода от сходства субъектов между собой в общности частных свойств и их несходства в плане общих свойств с неким третьим субъектом. В конце концов, предложена всего лишь рабочая модель, которая позволяет обратить внимание на методологическую необходимость использования формальных схем и обозначений для описания отношений искусственного и естественного типов интеллекта. Именно этот путь, путь символизации и абстрагирования открывает перспективы обнаружения методик понимания искусственного интеллекта как объекта, любое высказывание о котором не есть буквальная правда.

Центральное место в ряду методик, позволяющих представить искусственный интеллект как вид fiction, занимает семиотика. Причины выбора общей теории знаков и знаковых ситуаций в качестве методологической базы обусловлены теми богатыми возможностями, которые дает это направление для реализации фикционалистского понимания искусственного интеллекта. Наиболее важно здесь то, что семиотика позволяет затронуть искусственный интеллект как вид fiction в прагматическом аспекте. Речь, конечно, идет не о том, что семиотика – это лишь прагматика, в то время как формальная логика суть «чистая» семантика, а математика есть область «чистого» синтаксиса. Под «чистотой» понимается независимость от прочих вариантов работы со знаками и знаковыми ситуациями. Понятно, что утверждать, что семиотика, логика и математика обособили три взаимосвязанные сферы исследования знаков, будет в целом неверно. Однако речь идет о доминантах, обусловленных тем или иным типом знания. Конечно же, семантика и синтаксис рассматриваются в семиотике, которую Чарльз Пирс [10] представлял в рамках тесной связи с общей логикой. В том же отношении справедливо, что математика работает не только с числами, множествами, функциями и другими объектами, занимаясь исключительно отношениями математических знаков, равно как затрагивая правила работы с ними. Знаменитая исследовательская программа Давида Гилберта, как и некоторые изыскания Анри Пуанкаре [11], представили прекрасный пример семантических задач, которые также должны решать математики. Курт Гедель, предложив доказательство формальной неразрешимости оснований математики [12], тем самым внес вклад и в прагматические аспекты работы с математическими формализмами. И уж точно неверным будет утверждение, что формальная логика занимается исключительно способами связи знаков с предметами их обозначения. Исследователи в сфере логики по необходимости занимаются синтаксисом, а разработки Джона Остина в области теории речевых актов [13] дали на выходе варианты логической прагматики.

Автор признает условность высказанных идей. Причины, по которым нечто вообще утверждается о логике и математике, довольно просты: современные модели искусственного интеллекта имеют в основном вид математических алгоритмов и логических систем исчисления [14]. Именно в этом отношении важно, что семиотика способна показать, как в общем случае именно такие формализмы могут быть реализованы в сфере фикционалистского прочтения проблемы. Причем в контексте семиотики принципиально важным обстоятельством выступает привлечение концепции культуры как механизма, порождающего смыслы, предложенной Юрием Лотманом и Борисом Успенским [15]. Значительное число выдвинутых этими авторами идей, в частности: культура есть особый аппарат, работа которого исторически направлена на шифровку/дешифровку сообщений, нуждающихся в запоминании и трансляции, – применительно к вопросам искусственного интеллекта как варианта воплощения fiction позволяет отвлечься от строгости формальных построений, предложенных в логике и математике для моделирования поведения и общения. Открываются перспективы по означиванию культурных ситуаций, общее использование которого даст возможность сформировать основные принципы практической реализации искусственного интеллекта в рамках образования.

*2. Применение искусственного интеллекта в образовании*

Представление искусственного интеллекта в качестве литературной системы, имеющей семиотическое прочтение, тесно связано с культурной обусловленностью интеллектуальных действий как таковых. Не обязательно понимать интеллект или «дискурсивную силу мышления» буквально в виде текста, написанного некоторым автором для руководства к действию читателям, хотя и такое понимание тоже оправданно. Любые серьезные книги, а шире – художественные произведения: пьесы, кинофильмы, жи-вописные картины, музыкальные сочинения и другие творения, которые порождают в сознании своих читателей и/или зрителей непосредственный отклик, – все они могут быть поняты как формы искусственного интеллекта. Так, эстетика мира Средиземья, созданного Джоном Толкином в цикле известных произведений, была принята читателями буквально как руководство к действию. В итоге они стали участвовать в ролевых играх, воспроизводящих события из «Хоббита», «Властелина Колец» и других книг писателя. Аналогичным образом обстоит дело с вселенной «Звездных войн». Но еще более простым примером превращения художественного произведения в форму искусственного интеллекта является любая кинематографическая экранизация, влекущая за собой активизацию дискурсивной силы мышления режиссера, сценариста, актеров и других участников кинопостановки, которые в этот момент уподобляются организмам, фактически созданным воображением автора исходного произведения (книги, пьесы, сценария). Нетрудно увидеть, что автор в этом контексте играет роль «инопланетянина», интеллект которого в онтологическом плане сравнительно характеристик искусственного интеллекта, созданного человеком, был рассмотрен в первом разделе данной статьи.

Культурная обусловленность литературно-семиотической интерпретации искусственного интеллекта может быть продемонстрирована еще одним способом. Вслед за Юрием Лотманом и Борисом Успенским [15] будем полагать, что всякое сообщение в культуре представляет собой некий зашифрованный (закодированный) сигнал, требующий адресата, особой среды для своего формирования, каналов для передачи, памяти для хранения и т.д. В этом случае, например, образ Ивана Грозного, создаваемый в ходе интерпретации исторических памятников, в разные эпохи будет иметь различный вид (за точность примера авторы-основоположники ответственность не несут, ибо это пример автора данной статьи). Указанный образ будет представать то в виде олицетворения типичного тирана, то в качестве идеала мудрого правителя, борющегося за единство страны с внутренними и внешними ее врагами. И во всех случаях сообщения об Иване Грозном станут руководством к действию для представителей конкретной эпохи, расшифровываясь в рамках ожидаемых смыслов и преломляясь в сознании современников как структуре коллективной памяти.

Возможности реализации искусственного интеллекта в виде литературно-семиотической системы подразумевают выход на технологические приложения гуманитарных исследований. Этот момент тем более важен, что практически общепринятым в современности стало убеждение, что гуманитарные науки технологизации не подлежат, а если и подлежат, то лишь на платформе информационных технологий классического вида: компьютерных систем, программных продуктов, презентационных средств и других приспособлений. Именно такой путь видится для технологизации образования, в котором тенденцией последних лет стало применение так называемых «информационно-коммуникационных технологий» (ИКТ), причем применение безотносительно реальных нужд и возможностей образовательной деятельности. Привлечение информационных технологий дает возможность как минимум надеяться, что образовательная деятельность будет приведена к одному из стандартных видов:

(1) влияние факторов $A_1, A_2, A_3 \ldots A_n, A_{n+1}$ дает результаты $B_1, B_2, B_3 \ldots B_n, B_{n+1}$;

(2) факторы $A'_1, A'_2, A'_3 \ldots A'_n, A'_{n+1}$, дают результаты $B'_1, B'_2, B'_3 \ldots B'_n, B'_{n+1}$.

В то же время стандартизация, достигнутая в ходе применения ИКТ, открывает перспективы для симуляции образовательной деятельности, подмены образовательных действий квазиобразовательными операциями. Например, учитель истории, пользующийся средствами ИКТ, может апеллировать во время урока к графическим образам исторических событий, динамически меняющимся вариантам положения стран на политической карте и другим версиям представления учебной информации. Ученики получают информацию непосредственно на уроке, записывают ее на свои электронные устройства, сверяют с другими сведениями в глобальной информационно-коммуникационной сети «Интернет». В итоге обнаруживаются возможности по измерению скорости усвоения учебного материала, его объема и т.д. Но единиц для измерения степени усвоения информации (не объема, а именно степени!) указанная последовательность действий все-таки не предполагает. Техническое опосредование образовательных актов разрывает непосредственную связь учителя с учеником. В итоге учитель с трудом может знать, что же творится у его ученика в голове и что на самом деле, хорошо или не очень, усвоил ученик из всего объема полученной информации. Подлинная технологизация образования тем самым принципиально затрудняется. Проблема, однако, здесь видится в том, что причиной низкой технологизации гуманитарного знания выступает не природа самого этого знания, а достаточно узкое понимание любой технологии в качестве совокупности действий, приводящих к получению результатов по заданным параметрам. Расширительное толкование технологии позволит получить новые возможности в отношении гуманитарного знания и образования.

В естествознании имеются инвариантные единицы (килограммы, метры, секунды и другие), привлечение которых открывает перспективы для ожидания заданных результатов от конкретной последовательности учебных действий. В частности, при обучении спортсменов есть смысл полагаться на разработки в сфере медицины и строить методики на базе представлений, допустим, о работе сердечнососудистой системы. Такие методики способны сделать будущих спортсменов сильнее, быстрее, выносливее. Но гуманитарии работают с другими объектами, часть из которых вообще не поддается измерению, например уровень нравственного развития. Да, конечно, можно судить по косвенным признакам, насколько человек повел себя нравственно в той или иной ситуации. Но саму нравственность, в отличие от силы электрического тока или частоты сердечных сокращений, измерить довольно трудно. При расширительном толковании понятия «технология», предполагающем культурную обусловленность применений того или иного технологического решения, указанные трудности легко преодолеть. Технологией может быть названа последовательность действий, которая приводит к результатам, соразмерным горизонту ожиданий. При таком подходе мы не сможем строго сказать, быстрее спортсмен стал бегать или он окончательно лишился способности к прогрессу. И в этом смысле гуманитарное понимание технологии неактуально для естественнонаучных разработок. Но можно будет сказать, насколько некий объект изменился в целом, а главное, в каком направлении это произошло и почему другие направления реализованы быть не могли. Именно в этом случае становится возможной и измеримость нравственности, душевных качеств у спортсмена, его целеустремленности и силы воли.

Образование выступает средой, в которой некоторая культурная модель поведения, принятая определенными людьми, начина-ет генерировать для этих людей определенные смыслы, получая специфические формы и каналы трансляции. Вместе с тем литературный характер искусственного интеллекта, представленного в виде воплощения fiction и реализованного в образовании, предполагает в основе последовательность действий, которая ведет к складыванию ряда результатов в горизонте ожидания. Нетрудно увидеть, что отображение такой формы искусственного интеллекта включает как минимум два компонента, а именно методологическую культуру и функциональный характер словообразования.

Методологическая культура способствует росту навыков самостоятельного познания, представая в образовании в виде особого освоения учащимся изучаемого предмета. В рамках изучения физики это может быть представлено в виде навыка самостоятельной формулировки задач, решаемых в свете задействованной из различных источников теоретической информации. При изучении художественной литературы ту же роль могут играть тексты, освоенные в итоге их самостоятельного сравнения и интерпретации. Результатом роста методологической культуры является формирование дискурсивной силы мышления учащегося, направленной на поиск, систематизацию и обновление информации по изучаемому предмету. Достигаться это может, в том числе, при помощи средств ИКТ, которые выступают тем самым лишь инструментом в руках подготовленного пользователя, но не подменяют образовательный процесс как таковой.

Привлечение в контексте данного исследования функционального подхода к словообразованию, предложенного в лингвистике XX века [16: 3-4], может вызвать вопросы. Действительно, до сего момента речь шла о явлениях, которые к словам и словосочетаниям по форме не сводятся, а традиционно понимаются в качестве нематериальных воображаемых объектов. Тем самым требование опоры на методологическую культуру, которая не имеет однозначной референции, приобретает вид метафизического постулата. Фикционализм, в основе которого утверждения по типу «как если бы», в полной мере позволяет преодолеть эту трудность. Опора на лингвистические идеи, а именно: разделение языка на уровни (от морфемы до текста); определенность функций каждого из уровней на всяком более высоком, нежели предшествующий уровень; интерпретация языка как в целом коммуникативного аппарата – все это дает шанс представить процесс роста методологической культуры, как если бы у нее была однозначная референция. В этом случае методологическая культура растет или сокращается в системе координат «новое–старое» («инновационное–традиционное»). Всякое учебное действие, оформленное как производство новых слов, предложений и текстов, которые оппонируют сложившимся стереотипам в описании изучаемых ситуаций, есть признак роста. Верно и обратное: всякое действие, уменьшающее новизну используемых лингвистических средств и умножающее стереотипность описания, есть признак сокращения методологической культуры, а вместе с ней и дискурсивной силы мышления учащихся. Границы роста методологической культуры подвижны, а инновационность, как ее признак, относительна, причем корректируется нормами используемого языка, главная функция которого – обеспечить коммуникацию участников образовательного процесса.

**Заключение**

Интерпретация искусственного интеллекта с позиций философского фикционализма позволила обсудить условия технологизации образования как процесса по усилению уровня дискурсивной силы мышления учащихся. Искусственный интеллект получил расширительное истолкование как воплощение литературного образа, который не имеет однозначной связи с областью логических формализмов и математических алгоритмов действия и общения. Автор применил методы, свойственные семиотическому подходу, предложенному Борисом Успенским и Юрием Лотманом. Удалось подвергнуть критике современные версии технологизации образования, ведущие к приданию абсолютной ценности информационно-телекоммуникационным технологиям. Рост методологической культуры учащихся, оформленный в виде функционального словообразования новых средств описания изучаемых предметов, был связан с развитием дискурсивной силы мышления. Результаты исследования открыли новые перспективы в понимании места и роли искусственного интеллекта в рамках образовательной практики.

***Литература:***
1. *Vaihinger H.* The Philosophy of 'As if': A System of the Theoretical, Practical and Religious Fictions of Mankind. Eastford: Martino Fine Books, 2009.

2. *Salis F.* Fictional Reports. A Study on the Semantics of Fictional Names // Revista de Teoría, Historia y Fundamentos de la Ciencia. 2010. Vol. 25. No. 68. – P. 175–185.
3. *Соколов А.В.* Информатические опусы. Опус 5. Природа и сущность информации // Научные и технические библиотеки [Informatics opuses. Opus 5. Nature and essence of information]. 2011. No. 2. C. 5–27. (*Sokolov A.V.* Informaticheskie opusy. Opus 5. Priroda i sushchnost' informatsii // Nauchnye i tekhnicheskie biblioteki [Scientific and Technical Libraries]. 2011. No. 2. – P. 5–27.
4. *Steel S.* Recovering Ancient and Medieval Contemplative Taxonomies as an Alternative to Bloom's Taxonomy of Educational Objectives // Paideusis. 2012. Vol. 20. No. 2. – P. 46–56.
5. *Cronbach L.J.* The two disciplines of scientific psychology // American Psychologist. 1957. Vol. 12. – P. 671–684.
6. *Hunt E.* Human Intelligence. New York: Cambridge University Press. 2010. – 524 p.
7. *Hsu F.-H.* Behind Deep Blue – Building the Computer that Defeated the World Chess Champion. Princeton: Princeton University Press, 2004. – 320 p.
8. *Kripke S.* Naming and Necessity. Oxford: Basil Blackwell, 1980. – 172 p.
9. *Rorty R.* Philosophy and the Mirror of Nature. Princeton: Princeton University Press. 1979. – 401 p.
10. *Peirce C.S.* Necessity, The Doctrine of-Examined // Monist. 1891/1892. Vol. 2. – P. 321–337.
11. *Ewald W.B. (ed.)* From Kant to Hilbert: A Source Book in the Foundations of Mathematics: in two volumes. Oxford: Oxford University Press, 1996. – 650 p.
12. *Gödel K.* On Formally Undecidable Propositions of Principia Mathematica and Related Systems. New York: Dover Publications, Inc., 1992. – 72 p.
13. *Austin J.L.* How to do Things with Words. Second edition. Cambridge, MA: Harvard University Press, 1975. – 192 p.
14. *Kotthoff L.* Algorithm Selection for Combinatorial Search Problems: A Survey // AI Magazine. 2014. Vol. 35. No. 3. – P. 48–60.
15. *Лотман Ю.М., Успенский Б.А.* О семиотическом механизме культуры // Лотман Ю.М. Избранные статьи: в 3 т. Таллинн: Александра, 1992-93. Т. 3. – С. 326–344. (*Lotman J.M., Uspensky B.A.* About the Semiotics Mechanism of Culture. Izbrannye stat'i. Tallinn: Aleksandra Publ., 1992-93. Vol. 3. – P. 326–344.
16. *Резанова З.И.* Функциональный аспект словообразования: Русское производное имя. Томск: Изд-во Том. ун-та, 1996. – 219 с. (*Rezanova Z.I.* Functional Aspect of Word Formation: Russian Derivative Name. Tomsk: Tomsk State University Press, 1996. – 219 p.